\documentclass[12pt]{elsart}

\usepackage[T1]{fontenc}
\usepackage[utf8]{inputenc}
\usepackage[english]{babel}

\usepackage{cite,amsmath,amssymb,colordvi,graphicx}
\begin{document}

\begin{frontmatter}

\title{New nonlinear equation of KdV-type\\ for a non-flat bottom }

\author[AK]{Anna Karczewska}
\address[AK]{Faculty of Mathematics, Computer Science and Econometrics\\ University of Zielona G\'ora,
 ul.\ Szafrana 4a, 65-516 Zielona G\'ora, Poland} 
\ead{A.Karczewska@wmie.uz.zgora.pl} 

\author[PR]{Piotr Rozmej}
\ead{P.Rozmej@if.uz.zgora.pl}
\author[PR]{and Łukasz Rutkowski}
\address[PR]{ Faculty of Physics and Astronomy\\ 
 University of Zielona G\'ora, Szafrana 4a, 65-516 Zielona G\'ora, Poland}

\begin{abstract}
In the paper a new nonlinear equation describing shallow water waves with the topography of the bottom directly taken into account is derived.  This equation is valid in the weakly nonlinear, dispersive and long wavelength limit. Some examples of soliton motion for various bottom shapes obtained in numerical simulations according to the derived equation are presented.  
\end{abstract}


\end{frontmatter}

\section{Introduction}\label{Int}
In 1834 John Scott Russell observed the solitary wave.
His report to the British Association on that observation and further experiments in 1844 opened a new field in physics, concerning phenomena governed by nonlinear differential equations with soliton solutions. 
This type of equations appears in many fields,
from hydrodynamics, through electric currents, light propagation in fibres to some quantum phenomena (e.g.\ nonlinear Schr\"odinger equation). There are many journals specialized in nonlinear phenomena in which a lot of new results appear every year.   

One of the classical problems originating from J.S.~Russell's studies is a so called {\em shallow water wave problem}. The first single equation describing Russell's observation was obtained and solved by D.J.\ Korteveg and G. de Vries \cite{kdv} already in the nineteenth century. Their famous KdV equation became a prototype of exactly solvable soliton equations. The  theory of solitons can be found in several monographs, e.g., \cite{Ablc,Bel,Abl,DrJ,Rem,InR,Osb,Hir}. That deep mathematical theory is, however, limited to such cases of nonlinear wave equations which correspond to the flat bottom in the shallow water wave problem. In the past a lot of efforts were done in order to describe water wave motion over a bottom topography. Among them the Gardner equation sometimes called the forced KdV equation \cite{Grim,Grim1,Grim2} has been discussed extensively. Another approaches are the generalized Boussinesq equations \cite{Mit}, the Green-Naghdi equations \cite{GN,GN1,GN2} and an analytical and  numerical approach to the system of the Euler equations with appropriate boundary conditions \cite{Peli,Peli1}. 
It is impossible to list even a small fraction of the papers related to the field as they are so numerous.

Our approach is different. Assuming that the amplitude of the function describing the bottom is relatively small and its derivatives are not very big, either, we apply a perturbative approach to the system of hydrodynamic equations and derive entirely new nonlinear equations for surface waves in the weakly dispersive limit of shallow water problem.

The paper is organized as follows: in Section \ref{Prel} the system of equations for an ideal fluid is presented in dimensional and non-dimensional variables. The new small parameter $\delta$ related to the amplitude of the bottom variation is introduced and the set of two coupled equations (Boussinesq-type system) up to the second order in small parameters is obtained. The general method of elimination of one of the variables \cite{B&S} allowing to obtain a single differential equation for the wave is explained in details in Section \ref{SubAp}. 
For the flat bottom a generalized KdV equation, the second order in all small parameters obtained in \cite{B&S} is recovered. 
The new second order equation (in three small parameters) appropriate for a non-flat bottom is derived in Section~\ref{nflat}. In Section~\ref{nsim} several examples of time evolution of KdV soliton obtained with numerical simulations according to derived equations are presented and discussed.

%\section{Fluid equations}\label{Prel}

\section{Equations for irrotational, incompressible and inviscid fluid}\label{Prel}

In the standard approach to KdV equation the problem is considered as an irrotational motion of an inviscid incompressible fluid in a container with the flat bottom. Therefore a velocity potential $\phi$ is introduced, which fulfills the Laplace equation with appropriate boundary conditions. The system of equations for the velocity potential $\phi(x,y,z,t)$ can be found in many textbooks, for instance, see  \cite[Eqs.\ (5.2a-d)]{Rem}
\begin{eqnarray}  \label{g1}
\phi_{xx} + \phi_{yy} + \phi_{zz}&=& 0,  \\ \label{g2}
\phi_z - (\eta_x \phi_x + \eta_y \phi_y +\eta_t) &=& 0,  \\ \label{g3}
\phi_t + \frac{1}{2}(\phi_x^2+\phi_y^2+\phi_z^2) +g\eta &=& 0, \\ \label{g4}
\phi_z &=& 0, 
\end{eqnarray}
where eq.\ (\ref{g1}) is valid for $0<z<H+\eta(x,y,t)$, whereas eqs.\ (\ref{g2}-\ref{g3}) only at $z=H+\eta(x,y,t)$ and (\ref{g4}) at $z=0$. 

In the equations (\ref{g1}-\ref{g4}), $z=0$ at the flat bottom, $H$ is the level of unperturbed water surface, $g$ is the acceleration due to the gravity and $\eta(x,y,t)$ denotes the wave elevation with respect to $H$. For abbreviation all partial derivatives are denoted by appropriate subscript indexes through the whole paper (i.e., $\phi_{xx}\equiv\phi_{2x}\equiv \partial^2\phi/\partial x^2$, and so on).  The equations (\ref{g2}) and (\ref{g3}) represent so called kinematic and dynamic boundary conditions on the surface of the fluid, respectively. The equation (\ref{g4}) is the boundary condition at an impenetrable bottom (the component of water particle velocity perpedicular to the bottom has to vanish). By now, a very small surface tension term is neglected, but in general it can be taken into account.

 \begin{center}\begin{figure}[htb]\resizebox{0.8\columnwidth}{!}{\includegraphics{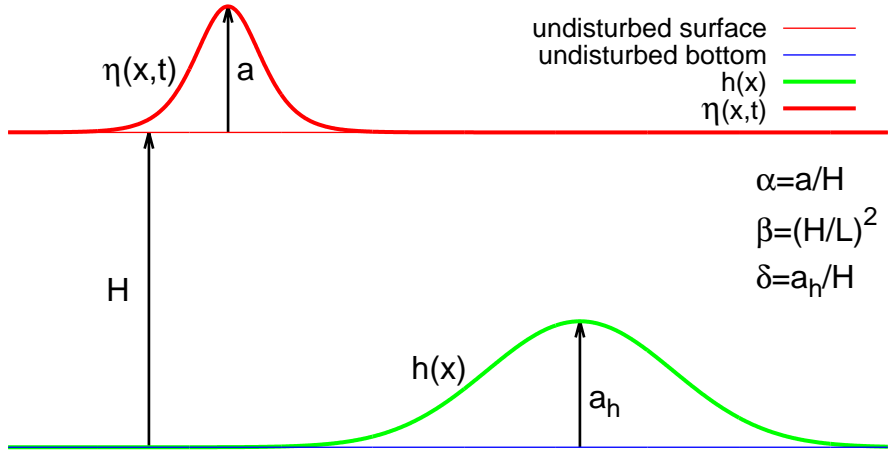}}\caption{Schematic view of the geometry of shallow water wave problem in two dimensions.} \label{F0}\end{figure}\end{center}
In the following we limit our considerations to the 2-dimensional flow, $\phi(x,z,t), \eta(x,t)$, where $x$ is the horizonatal coordinate and $z$ is the vertical one (it means translational symmetry with respect to $y$ axis). In the same spirit like Burde and Sergyeyev \cite{B&S} do we introduce two small non-dimensional parameters which can control the order of approximation. Then, $\alpha=a/H$, where $a$ is an amplitude of surface wave $\eta$ and $\beta=H^2/L^2$, where $L$ is a typical wavelengths of surface waves. The parameters  $\alpha,\beta$ are the same as the parameters $\varepsilon,\delta^2$ in \cite{Rem}, respectively. With those parameters the non-dimensional variables are defined as follows 
\begin{equation} \label{bezw}
\left. \begin{array}{lll}
&\quad \tilde{\eta}=\eta/a,& \tilde{\phi}= \phi /(L\frac{a}{H}\sqrt{gH}),\\  \tilde{x}= x/L, & \quad \tilde{z}= z/H, & \quad \tilde{t}= t/(L/\sqrt{gH}). 
\end{array} \right.
\end{equation}  
In the non-dimensional variables the equations  (\ref{g1}-\ref{g3}) for 2-dimensional flow take the form (below all tildes have been omitted)
\begin{eqnarray} \label{2BS}
\beta \phi_{xx}+\phi_{zz} &=&0,   \\ \label{4BS}
\eta_t+\alpha\phi_x\eta_x-\frac{1}{\beta}\phi_z &=&0,  \\ 
 \label{5BS}
\phi_t+\frac{1}{2}\alpha \phi_x^2+\frac{1}{2}\frac{\alpha}{\beta}\phi_z^2 +\eta &=& 0 .
\end{eqnarray}
Like above,  eq.\ (\ref{2BS}) is valid for $0\le z \le 1+\alpha \eta$ and
eqs.\ (\ref{4BS}-\ref{5BS}) only at $z = 1+\alpha \eta$.
Our main goal is to introduce a non-flat bottom in the general form $h(x)$ in the original (dimensional) variables. For that case the boundary condition (\ref{g4}) has to be replaced by 
\begin{equation} \label{2org}
\phi_z=h_x\,\phi_x.
\end{equation}
In non-dimensional variables $\tilde{h}= h/a_h,$ where $a_h$  is the amplitude of the bottom variation with respect to the flat part. Then introducing the parameter $\delta=a_h/H$ we obtain the boundary condition at the bottom in the form (again omitting tildes)
\begin{equation} \label{2nn}
\phi_z=\beta\delta\left( h_x\,\phi_x\right) \quad \mbox{for} \quad z=\delta\, h(x) 
\end{equation}
i.e., at the general bottom function.
In order to assure that the perturbative approach make sense we have to assume that derivatives of $h(x)$ are everywhere small enough. 

We search for the solution of the equations  (\ref{2BS}-\ref{5BS},\ref{2nn}) in the standard form of series
\begin{equation} \label{sz1}
\phi(x,z,t)=\sum_{m=0}^\infty z^m\, \phi^{(m)} (x,t).
\end{equation}
The equation (\ref{2BS}) imposes the following conditions on the functions ~$\phi^{(m)}(x,t)$ 
\begin{equation} \label{Lrec}
\begin{array}{llll}
\phi^{(2k)} &  = &  \frac{(-\beta)^m}{(2m)!}\,\phi^{(0)}_{2kx} & \quad \mbox{for~~$m=2k$~~even}\\
\phi^{(2k+1)} & =& \frac{(-\beta)^m}{(2m)!}\,\phi^{(1)}_{2kx} & \quad \mbox{for~~$m=2k+1$~~odd}
\end{array}
\end{equation}
which allow to express all those functions by $\phi^{(0)},\phi^{(1)} $ and their derivatives with respect to $x$. For the lowest terms of expansion (\ref{sz1}), $m\le 6$ the relations  (\ref{Lrec}) read as
\begin{equation} \label{recur}
\left. \begin{array}{lll}
\phi^{(2)}=-\frac{1}{2}\beta\phi^{(0)}_{2x}, &
\phi^{(3)}=-\frac{1}{6}\beta\phi^{(1)}_{2x},&  \phi^{(4)}=\frac{1}{24}\beta^2\phi^{(0)}_{4x},\\
 \phi^{(5)}=\frac{1}{120}\beta^2\phi^{(1)}_{4x},&  \phi^{(6)}=-\frac{1}{720}\beta^3\phi^{(0)}_{6x}. & 
\end{array} \right. 
\end{equation}
Then the boundary condition at the bottom (\ref{2nn}), with conditions (\ref{Lrec})  takes form
\begin{eqnarray} \label{ward}
0& = & \phi^{(1)} 
- \beta\delta\left( h_x \phi^{(0)}_x + h \phi^{(0)}_{2x}\right)- \beta\delta^2\left(h h_x \phi^{(1)}_x + \frac{1}{2} h^2  \phi^{(1)}_{2x} \right)  \\
&& \hspace{4.5ex}
+\beta^2\delta^3\left(\frac{1}{2}  h^2 h_x  \phi^{(0)}_{3x} 
+\frac{1}{6}  h^3\phi^{(0)}_{4x} \right)
 + \beta^2\delta^4\left(\frac{1}{6}  h^3 h_x  \phi^{(1)}_{3x}
+\frac{1}{24}  h^4\phi^{(1)}_{4x} \right)\nonumber .
\end{eqnarray}
It is easy to see that for $h=\mbox{const}$, ($\delta=0$), $\phi^{(1)}=0$ and all odd components  $\phi^{(2k+1)}=0$.

Terms of orders higher than $\beta^2\delta^4$ are omitted in (\ref{ward}). In general, the equation (\ref{ward}) should be solved to supply $\phi^{(1)}(x,t)$ as function of $\phi^{(0)}, h $ and their derivatives. It is simple only if one takes the boundary condition (\ref{ward}) in the lowest order in small parameters, $\beta\delta$
\begin{equation} \label{ord1}
\phi^{(1)}(x,t)= \beta\delta \left(h_x \phi^{(0)}_{x} +h  \phi^{(0)}_{2x}\right).
\end{equation}
In that approximation all functions $\phi^{(m)}$ can be expressed by the functions $\phi^{(0)}, h $ and their derivatives. Then the velocity potential is (limiting the series (\ref{sz1}) to $m=6$)
\begin{eqnarray} \label{pot1}
\phi(x,z,t)  & = &   \phi^{(0)}_{x}+z\beta\delta \left(h  \phi^{(0)}_x\right)_{x}
-\frac{1}{2}z^2 \beta \, \phi^{(0)}_{2x} 
  -\frac{1}{6}z^3 \beta^2\delta \left(h\phi^{(0)}_x \right)_{3x}  \\ &&
 + \frac{1}{24}z^4 \beta^2 \phi^{(0)}_{4x} 
 + \frac{1}{120}z^5 \beta^3\delta\left(h \phi^{(0)}_x \right)_{5x} 
 +  \frac{1}{720}z^6 \beta^3 \phi^{(0)}_{6x}. \nonumber
\end{eqnarray}
Next, we insert $\phi(x,z,t)$ given by (\ref{pot1}) into (\ref{4BS}) and (\ref{5BS}) neglecting terms of high order in small parameters $\alpha, \beta, \delta$. Then the latter equation is differentiated with respect to $x$. After the substitution $\phi^{(0)}_{x}(x,t)\equiv w(x,t)$ a system of coupled differential equations for $\eta(x,t)$ and $w(x,t)$ is obtained which can be considered at different orders of approximations.

Retaining only the terms of the second order in small parameters that system of equations becomes the second order Boussinesq system
\begin{eqnarray} \label{4hx}
\eta_t + w_x  & +&    \alpha\,(\eta w)_x-\frac{1}{6}\beta\, w_{3x} - \frac{1}{2} \alpha\beta \,(\eta w_{2x})_x +\frac{1}{120}\beta^2\, w_{5x} \nonumber \\  & -&
 \delta\, (hw)_x +\frac{1}{2}\beta\delta \,(hw)_{3x}=0  
\end{eqnarray}
and
\begin{eqnarray} \label{5hx}
w_t  & +& \eta_x + \alpha\, w w_x-\frac{1}{2}\beta\, w_{2xt} +\frac{1}{24}\beta^2\, w_{4xt}  \\
   & +&
\frac{1}{2} \alpha\beta\left[-2(\eta w_{xt})_x +  w_x w_{2x}- w w_{3x} \right] + \beta\delta\, (h w_t)_{2x}  = 0 . \nonumber
\end{eqnarray}

\section{Subsequent approximations} \label{SubAp}

The method of subsequent approximations used by Burde and Sergyeyev \cite{B&S}
consists in using special properties of
solutions to lower order equations for $w$ and $\eta$ in derivations of corrections to solutions in the next order. 
First, we demonstrate the method by obtaining the equation for $\eta(x,t)$ up to the second order in parameters $\alpha$ and $\beta$, i.e., eliminating $w(x,t)$ from equations (\ref{4hx}-\ref{5hx}) with terms containing $\delta$ omitted (omitting terms with $\delta$ is equivalent to the assumption that $h=h(x)$ does not depend on $x$).

The equations for  $w$ and $\eta$ in the first order in  $\alpha$ and $\beta$ are well known, as they, after suitable change of variables, lead to standard KdV equation, see, e.g., \cite[Eq.\ (16)]{B&S} in which $\beta=\alpha$
\begin{equation} \label{16BS}
w = \eta +\alpha (-\frac{1}{4}\,\eta^2) +\beta\frac{1}{3}\,\eta_{2x}, \quad 
\eta_t + \eta_x + \alpha\frac{3}{2}\eta \eta_x+\beta\frac{1}{6} \eta_{3x}=0.
\end{equation} 

In order to find the second order equation (in $\alpha$ and $\beta$) for $\eta$ we postulate $w$ in the form
\begin{equation} \label{wab2}
w(x,t)  =  \eta-\frac{1}{4}\alpha\,\eta^2 +\frac{1}{3}\beta\,\eta_{2x} 
 + 
\alpha^2\, \mbox{Qa2}(x,t) +\beta^2 \, \mbox{Qb2}(x,t) +\alpha\beta\, \mbox{Qab}(x,t),
\end{equation}
where $\mbox{Qa2}(x,t), \mbox{Qb2}(x,t) \mbox{~and~} \mbox{Qab}(x,t)$ are unknown functions of $\eta$ and its derivatives. Then we insert the above form of $w$ into equations (\ref{4hx}-\ref{5hx}) in which the time derivatives of $\eta$ are replaced by space derivatives according to lower order approximation (\ref{16BS}), i.e. 
$$
\eta_t = -\eta_x - \alpha\frac{3}{2}\eta \eta_x-\beta\frac{1}{6} \eta_{3x}.$$
Rejecting terms of the order higher than the second in $\alpha$ and $\beta$ one obtains
the system of equations (in which the terms of the lower orders have already cancelled)
\begin{equation} \label{58a}
\alpha ^2 \left(\mbox{Qa2}_x-\frac{3}{4} \eta ^2 \eta_x\right)
+\beta ^2 \left(\mbox{Qb2}_x-\frac{17}{360} \eta_{5x}\right)
+\alpha  \beta    \left(\mbox{Qab}_x+\frac{1}{12} \eta_x
   \eta_{2x}-\frac{1}{12} \eta \eta_{3x}\right) =0, 
\end{equation} \vspace{-1ex}
\begin{equation}  \label{59a}
\alpha ^2 \mbox{Qa2}_t +\beta ^2 \left(\mbox{Qb2}_t+\frac{11}{72} \eta_{5x}\right) 
+\alpha  \beta    \left(\mbox{Qab}_t+\frac{11}{6} \eta_x
   \eta_{2x}+\frac{11}{12} \eta \eta_{3x}\right) = 0.
\end{equation}
Because $\alpha$ and $\beta$ are independent parameters we can subtract those equations and require the vanishing of each coefficient standing at $\alpha^2$, $\beta^2$ and $\alpha\beta$ separately. Then we obtain three equations
\begin{eqnarray} \label{roznice}
-\mbox{Qa2}_t+\mbox{Qa2}_x-\frac{3}{4} \eta ^2 \eta_x & = & 0, \\ \label{r1}
-\mbox{Qb2}_t +\mbox{Qb2}_x-\frac{1}{5} \eta_{5x}& = & 0, \\ \label{r2}
-\mbox{Qab}_t +\mbox{Qab}_t-\frac{7}{4} \eta_x \eta_{2x}- \eta \eta_{3x}& = & 0.
\end{eqnarray}
Like in the previous order of approximation we can substitute
$$\mbox{Qa2}_t=-\mbox{Qa2}_x,\quad \mbox{Qb2}_t=-\mbox{Qb2}_x,\quad \mbox{Qab}_t=-\mbox{Qab}_x .$$
Then each of equations (\ref{roznice}-\ref{r2}) can be integrated with respect to $x$ resulting in
\begin{equation} \label{Qab} 
\mbox{Qa2} = \frac{1}{8} \eta^3,~~ \mbox{Qb2} = \frac{1}{10} \eta_{4x},~~  \mbox{Qab} = \frac{3}{16} \eta_{x}^2 + \frac{1}{2}\eta \eta_{2x}  .
\end{equation}
Finally, we obtain 
\begin{equation} \label{wab2a}
w(x,t) = \eta-\frac{1}{4}\alpha\,\eta^2 +\frac{1}{3}\beta\,\eta_{2x}+\alpha^2\,\frac{1}{8} \eta^3+
\alpha\beta \left( \frac{3}{16} \eta_{x}^2 + \frac{1}{2}\eta \eta_{2x}\right) +\beta^2 \, \frac{1}{10} \eta_{4x} 
\end{equation}
and 
\begin{equation} \label{eta2}
\eta_t+\eta_x+ \alpha\, \frac{3}{2}\eta\eta_x +\beta\,\frac{1}{6}\eta_{3x} 
+ \alpha^2\,\left(-\frac{3}{8}\eta^2\eta_x\right) 
+ \alpha\beta\,\left(\frac{23}{24}\eta_x\eta_{2x}+\frac{5}{12}\eta\eta_{3x} \right)+\beta^2\,\frac{19}{360}\eta_{5x}=0. 
\end{equation}
The equations (\ref{wab2a}) i (\ref{eta2}), for ~$\beta=\alpha$, become identical with equations (20) and (21) from the paper \cite{B&S}. As Burde and Sergyeyev have shown, the same procedure, when applied for higher order approximations, leads to several known forms of KdV-type equations of higher order derived earlier by other authors in a more heuristic manner. However, that systematic treatment of small parameters allows to avoid some small errors in the previous derivations.

\section{Variable bottom terms} \label{nflat}

Now, we are in good position to include terms containing influence of a non-flat  bottom. These are the terms of the  first order  $\delta$  and the second order $\beta\delta$ in small parameters in (\ref{4hx}) and the first order $\frac{\beta\delta}{\alpha}$ and the second order $\beta\delta$ in (\ref{5hx}).

 However, the absence of the first order term with $\delta$ in the equation (\ref{5hx}) makes the derivation of the correction function of that order impossible. In other words the equations obtained for correction term in $w$ of the form $\delta\, \mbox{Qd}(x,t)$  become contradictory in the first order approximation. Therefore we make use of the following trick justified by the fact that the relation of $\phi_z$ to $\phi_x$ at the bottom is of the order of  $\beta\delta$, see (\ref{2nn}).
We include the term $-\delta (hw)_x$, contained in the equation (\ref{4hx}) into the second order term with $\beta\delta$
\begin{equation} \label{trick}
-\delta (hw)_x+\frac{1}{2}\beta\delta\ (hw)_{3x}
=\beta\delta \left(-\frac{1}{b} (hw)_x +  \frac{1}{2} (hw)_{3x}  \right).
\end{equation}

For derivation of the second order equations we keep $b$ as a constant different than $\beta$ but in the resulting equations we will put $b=\beta$.  
Then the system of equations  (\ref{4hx}) and (\ref{5hx}) can be written as
\begin{eqnarray}\label{4hxd}
\eta_t  + w_x &+& \alpha(\eta w)_x-\frac{1}{6}\beta w_{3x} -\frac{1}{2} \alpha\beta (\eta w_{2x})_x +\frac{1}{120}\beta^2 w_{5x}  \nonumber\\
&+& \beta\delta \left(-\frac{1}{b} (hw)_x + \frac{1}{2} (hw)_{3x} \right) =0 ,
\end{eqnarray}
and
\begin{eqnarray} \label{5hxd}
w_t &+& \eta_x + \alpha\, w w_x-\frac{1}{2}\beta\, w_{2xt}  +\frac{1}{24}\beta^2\, w_{4xt} + \beta\delta \,(h w_t)_{2x}  \nonumber
\\  &+&
 \alpha\beta\left(-(\eta w_{xt})_x + \frac{1}{2} w_x w_{2x}-\frac{1}{2} w w_{3x} \right)  =0 .
\end{eqnarray}
We can now use the results (\ref{wab2a}) and (\ref{eta2}) and postulate 
\begin{equation}\label{wab2b}
w(x,t) =  \eta-\frac{1}{4}\alpha\,\eta^2 +\frac{1}{3}\beta\,\eta_{2x}+\alpha^2\,\frac{1}{8} \eta^3 +\beta^2 \, \frac{1}{10} \eta_{4x} +
 \alpha\beta \left( \frac{3}{16} \eta_{x}^2 + \frac{1}{2}\eta \eta_{2x}\right) +\beta\delta \mbox{Qbd}(x,t).
\end{equation}

Then we insert (\ref{wab2b}) into (\ref{4hx}) and replace the time derivatives of $\eta$ by the expressions from (\ref{eta2}) retaining only terms up to the second order in $\alpha,\beta,\delta$. The resulting differential equation for $\mbox{Qbd}(x,t)$ can be integrated to obtain
% Insertion of (\ref{wab2b}) into (\ref{4hx}), replacement of time derivatives of $\eta$ by expressions from (\ref{eta2}) and neglection of terms of higher order than the second one leads to differential equation for $\mbox{Qbd}(x,t)$ which can be integrated to obtain
\begin{equation} \label{Qbd}
\mbox{Qbd}(x,t) = \frac{(2h-bh_{2x})\eta}{4b} - h_x\eta_x - \frac{3}{4}h\eta_{2x}.
\end{equation}

 \begin{center}\begin{figure}[tb] \resizebox{0.8\columnwidth}{!}{\includegraphics{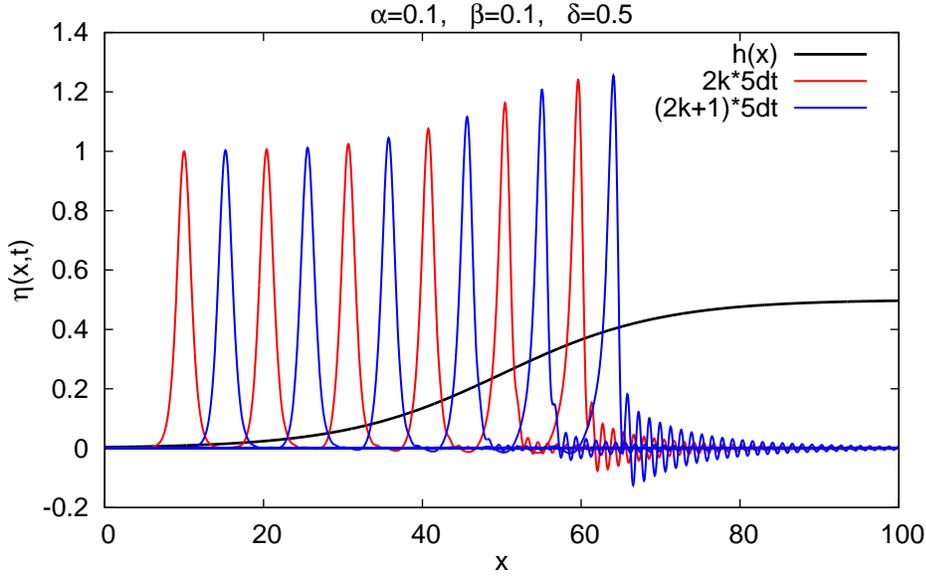}} \caption{Time evolution of the KdV soliton according to the equation (\ref{etaabd}) for decreasing water depth.} \label{F1} \end{figure}\end{center}

So, finally we obtain the equations (restoring $b=\beta$)
\begin{eqnarray} \label{wwabd}
w(x,t) &=& \eta-\frac{1}{4}\alpha\,\eta^2 +\frac{1}{3}\beta\,\eta_{2x}+\alpha^2\,\frac{1}{8} \eta^3 +\beta^2 \, \frac{1}{10} \eta_{4x}
+ \alpha\beta \left( \frac{3}{16} \eta_{x}^2 + \frac{1}{2}\eta \eta_{2x}\right) \\
&& +\beta\delta\left( \frac{(2h-bh_{2x})\eta}{4\beta} - h_x\eta_x - \frac{3}{4}h\eta_{2x}\right)\nonumber
\end{eqnarray}
and
\begin{eqnarray} \label{etaabd}
\eta_t+\eta_x &+& \alpha\, \frac{3}{2}\eta\eta_x +\beta\,\frac{1}{6} \eta_{3x} + \alpha^2\,\left(-\frac{3}{8}\eta^2\eta_x\right) +
 \alpha\beta\,\left(\frac{23}{24}\eta_x\eta_{2x}+\frac{5}{12}\eta\eta_{3x} \right)+\beta^2\,\frac{19}{360}\eta_{5x} \nonumber \\
 &+&\beta\delta\left(-\frac{1}{2}(h\eta)_x /\beta +\frac{1}{4} (h_{2x}\eta)_x -\frac{1}{4}(h\eta_{2x})_x \right) =0.
\nonumber
\end{eqnarray}
However, the terms containing $\beta$ in denominators have to be considered as belonging to the second order approximation.

The equation (\ref{etaabd}) is the first KdV-type equation containing the influence of the bottom topography in the lowest order. It is not yet clear whether analytical solutions of  (\ref{etaabd}) for some cases of the bottom function $h(x)$ can be found. One can try the famous {\em Inverse Scattering Transform} (IST) method \cite{GGKM,Ablc}, but its application can be very difficult.
On the other hand, numerical solutions for some particular initial conditions can be obtained relatively simply and they may inspire analytical studies, as happened in the past for the KdV case \cite{ZK,GGKM}.

The higher order approximations with respect to the bottom topography are, in principle, possible by taking also into account the term of the order $\beta\delta^2$ in the boundary condition at the bottom (\ref{ward}). However, that step introduces many more complications because  $\phi^{(1)}$ is then determined by the second order differential equation 
\begin{equation} \label{ward2}
\phi^{(1)} + \beta\delta (-h_x \phi^{(0)}_{x} -h  \phi^{(0)}_{2x}) +\beta\delta^2(-h h_x\phi^{(1)}_{x}-\frac{1}{2}h^2\phi^{(1)}_{2x})=0.
\end{equation}

Therefore, that problem will be discussed in the subsequent paper.

\section{Numerical simulations} \label{nsim}

\begin{center}\begin{figure}[bt] \resizebox{0.8\columnwidth}{!}{\includegraphics{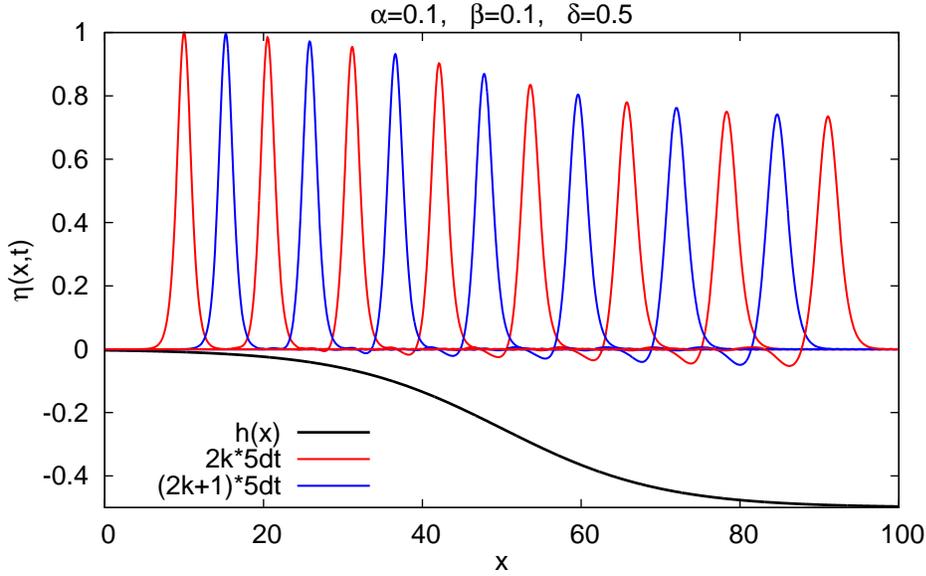}} \caption{Time evolution of the KdV soliton %according to the equation (\ref{etaabd}) 
for $\alpha=\beta=0.1 ~\delta=0.5$ for increasing water depth.} \label{F2} \end{figure}\end{center}

In this section some examples of numerical calculations of time evolution according to the second order equations (\ref{eta2}) and (\ref{etaabd}) are presented and discussed. In calculations the Zabusky-Kruskal algorithm \cite{ZK} was applied,  modified in order to compute space derivatives of $\eta(x,t)$ with high accuracy. All  calculations are done in non-dimensional variables (\ref{bezw}). The initial condition was always taken in the form of the exact KdV soliton (the solution of first order KdV equation  (\ref{16BS})). 
 Calculations were performed on the interval $x\in[0,X]$ with periodic boundary conditions. The space step in the grid was chosen to be $\Delta x=0.05$ and the time step $\Delta t=(\Delta x)^3/4$, like in  \cite{ZK}. In all simulations the volume of the fluid was conserved up to 10-11 digits. 

%\subsection{Results for the second order equation with flat bottom}

First we calculated the time evolution of the exact KdV soliton according to second order equation  (\ref{eta2}). When $\alpha=\beta=0.1$ the soliton moves almost unchanged for a long time.  That behaviour persists even for larger values of small parameters ($\alpha=\beta=0.15$) though distortions of the tails of the soliton, very small in the previous case, become a little bigger.

 In the case $\alpha >\beta$ one obtains that nonlinear terms prevail and the initial soliton quickly evolves into two-soliton solution.
In the opposite case $\alpha <\beta$ dispersive terms predominate and one observes decreasing amplitude of the wave, increasing width, distortion of the shape and creation of the wave trains at the tails.
All these effects are known from the analysis of the pure KdV equation (\ref{16BS}). Up to reasonable values of small parameters, $\alpha, \beta \lessapprox 0.15$, the same behaviour is preserved for solutions to the second order equation (\ref{eta2}).

Below several preliminary results for the time evolution of solutions to the second order equation with bottom topography included (\ref{etaabd}) are presented. 
The simulations show the influence of $h(x)$-dependent terms. 
For the beginning let us present the case of smooth decreasing (or increasing) of the water depth. The cases are shown in Figs.~\ref{F1} and \ref{F2}, respectively. The  non-dimensional $h(x)$ function was chosen in the form $h(x)=\pm \frac{1}{2}(\mbox{tanh}(0.05(x-50))+1)$. 
In both cases the calculations were performed on the interval $x\in [0,200]$ with $N=4000$ grid points, where the bottom function for the subinterval  $x\in [100,200]$ was symmetric to that in the subinterval $x\in [0,100]$. Such setting assured almost exact smoothness of the function $h(x)$ and its derivatives. The cases shown in Figs.~\ref{F1} and \ref{F2} model incoming and outgoing sea-shore waves. 
\begin{center}\begin{figure}[t]
\resizebox{0.8\columnwidth}{!}{\includegraphics{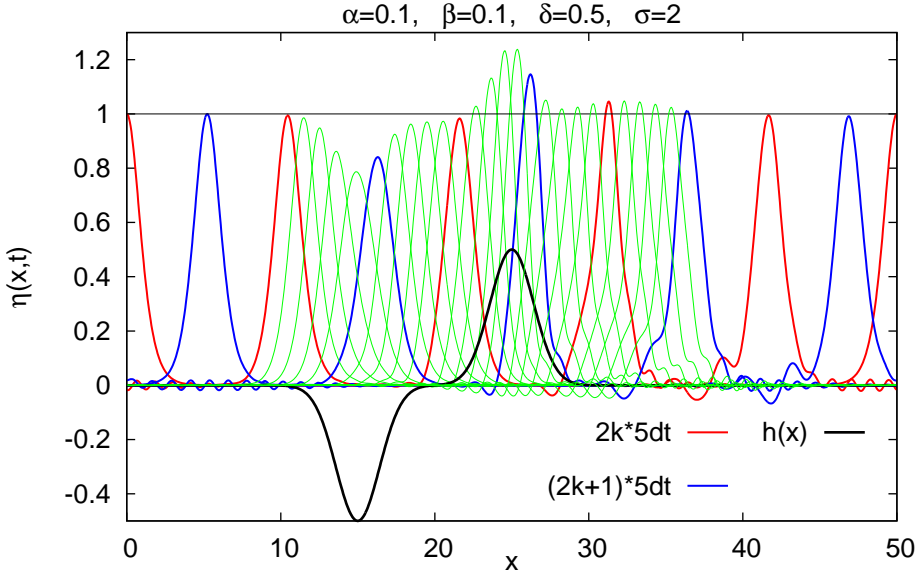}}
\caption{Time evolution of the KdV soliton according for the Gaussian well followed by the symmetric hump.} \label{F3}
\end{figure}\end{center}

\begin{center}\begin{figure}[t]
\resizebox{0.8\columnwidth}{!}{\includegraphics{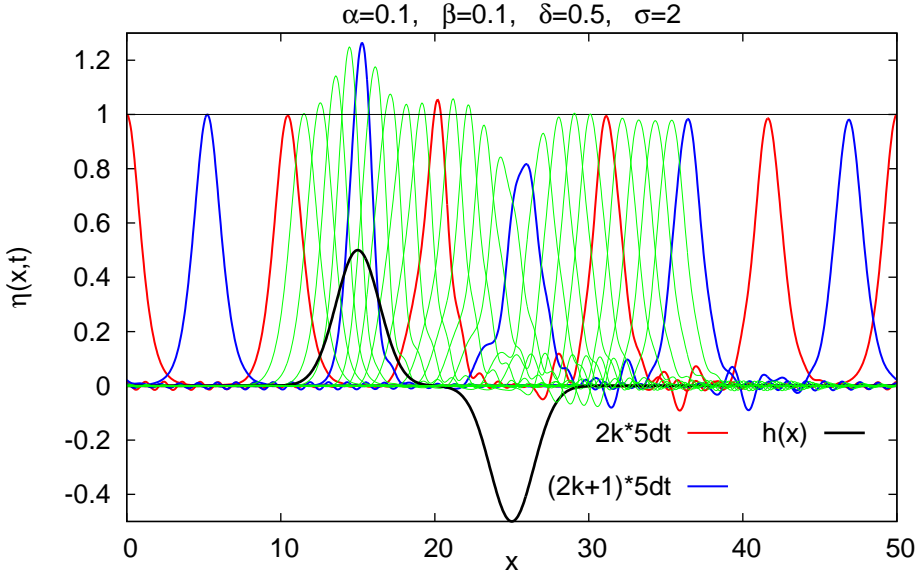}}
\caption{Time evolution of the KdV soliton for the Gaussian hump followed the well.} \label{F4}
\end{figure}\end{center}

In the figures \ref{F1}-\ref{F3} the bottom function is drawn with the thick black line. Medium thick blue lines represent the wave shapes at $t=0,10,20,\ldots$ and the red ones at $t=5,15,25,\ldots$
In Fig.~\ref{F1} we see a growth of the wave amplitude, forward scattering and formation of the shock wave when the soliton approaches the shallow region. In Fig.~\ref{F2} the wave slows down and decreases its amplitude when the water depth increases. At the same time a backward scattered wave appears.

\begin{center}\begin{figure}[t]
\resizebox{0.8\columnwidth}{!}{\includegraphics{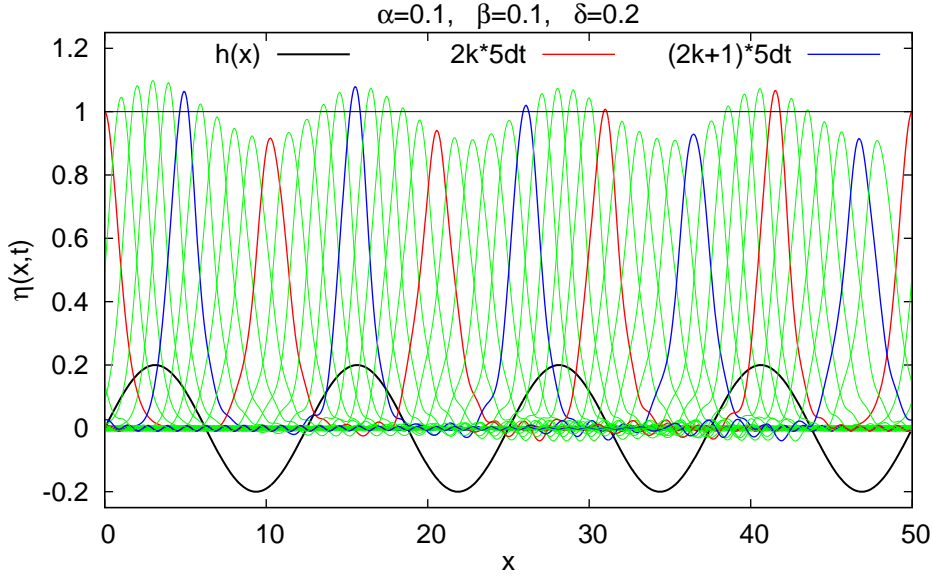}}
\caption{Motion of the soliton according to the equation (\ref{etaabd}) for periodically varying bottom.} \label{F6}
\end{figure}\end{center}

Figs.~\ref{F3} and \ref{F4} show the soliton motion for the bottom containing a well and a hump. In both cases $\alpha=\beta=0.1$ and $\delta=0.5$. The  bottom function was chosen as a sum of two Gaussians centered at $x=15$ and $x=25$ with widths $\;\sigma=2$. 
Calculations were performed on the interval $X\in [0:50]$ with 
$N=1000$  grid points and periodic boundary conditions.   In order to show details of the evolution thin green lines are plotted for $t\in [10,30]$ with the step $\Delta \,t=1$. In both cases we see decreasing (increasing) amplitudes of the wave passing over the well (hump), respectively. However, when the wave comes back to the flat part of the bottom it almost comes back to its original shape. 
The shape of the soliton evolving with respect to the equation (\ref{etaabd})
is resistant to bottom variations extended on long distances. The case presented
in Fig.~\ref{F6} shows the time evolution of the KdV soliton when the bottom function is the sinus function with the period $2\pi/12.5$ and the amplitude $0.2*H$. The soliton wave moves with its shape almost unchanged modifying its amplitude, width and speed. 
This behaviour maintains for larger values of $\delta$, only the distortions of tails become larger. 

\begin{center}\begin{figure}[tbh]
\resizebox{0.8\columnwidth}{!}{\includegraphics{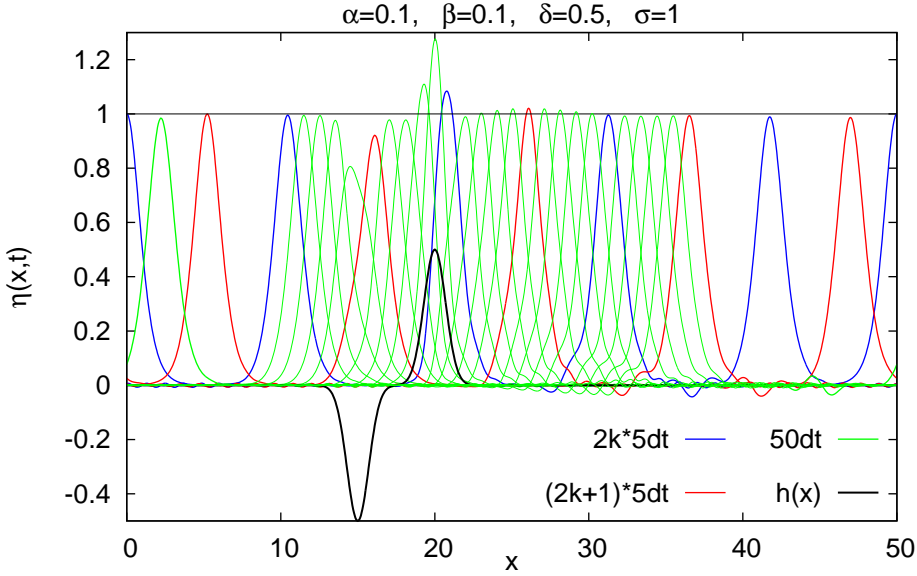}}
\caption{Motion of the soliton according to the equation (\ref{etaabd}) for a hole and a bump.} \label{F7}
\end{figure}\end{center}

\begin{center}\begin{figure}[tbh]
\resizebox{0.8\columnwidth}{!}{\includegraphics{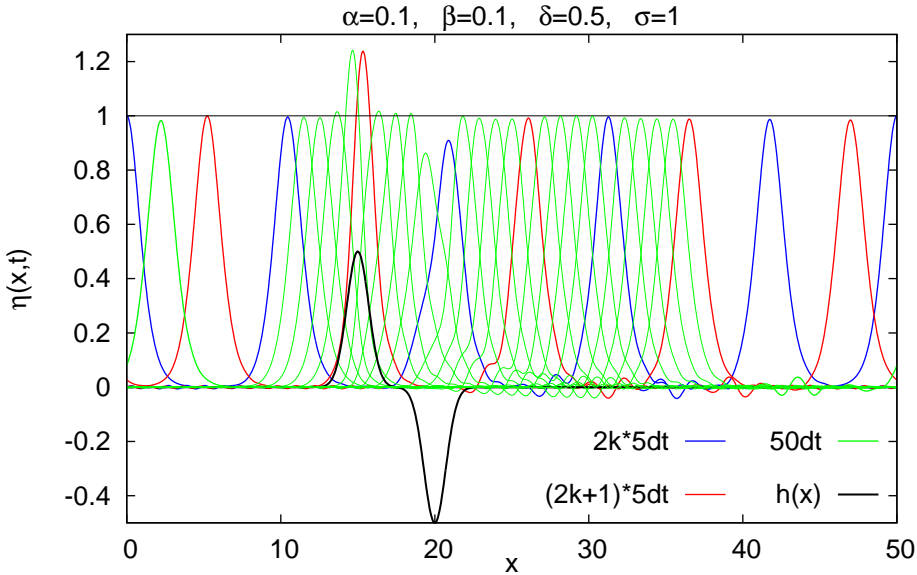}}
\caption{Motion of the soliton according to the equation (\ref{etaabd}) for a bump and a hole.} \label{F8}
\end{figure}\end{center}

In Figs.\ \ref{F7} and \ref{F8} cases similar to those from Figs.\ \ref{F3} and \ref{F4} are presented. This time the widths of the Gaussian wells and bumps  ($\sigma=1$) as well as the separation of the extrema are two times smaller.  Comparing  Figs.\ \ref{F7} and \ref{F8} to Figs.\ \ref{F3} and \ref{F4} one can notice the following differnces. In the former cases, where the interval of the obstacles is two times wider, the distortions of the soliton's shape are larger than in the latter cases. The amplitudes of back scattered waves are larger, too and the main soliton rquires a longer time period to regain its shape on the further motion over the flat bottom.

\begin{center}\begin{figure}[tbh]
\resizebox{0.8\columnwidth}{!}{\includegraphics{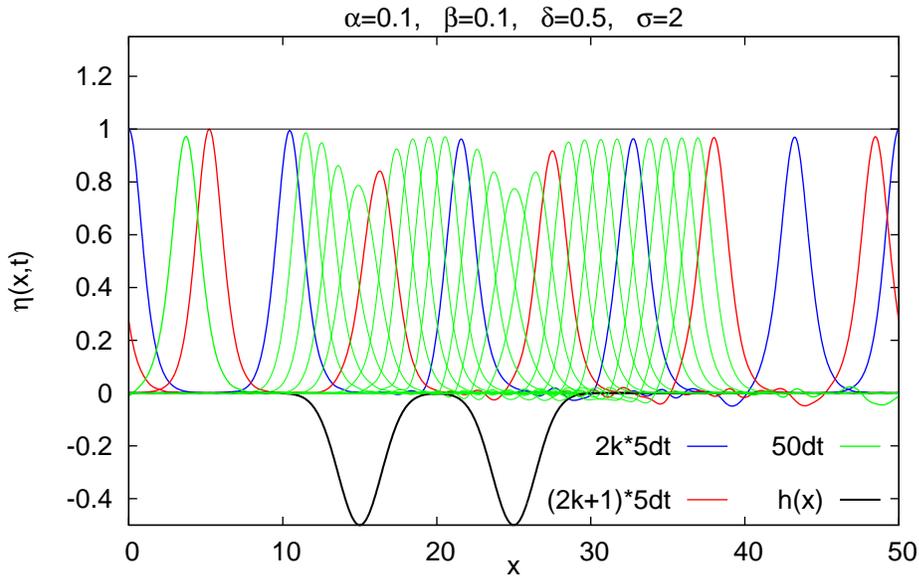}}
\caption{Motion of the soliton according to the equation (\ref{etaabd}) for a sequence of two wells.} \label{F9}
\end{figure}\end{center}

\begin{center}\begin{figure}[tbh]
\resizebox{0.8\columnwidth}{!}{\includegraphics{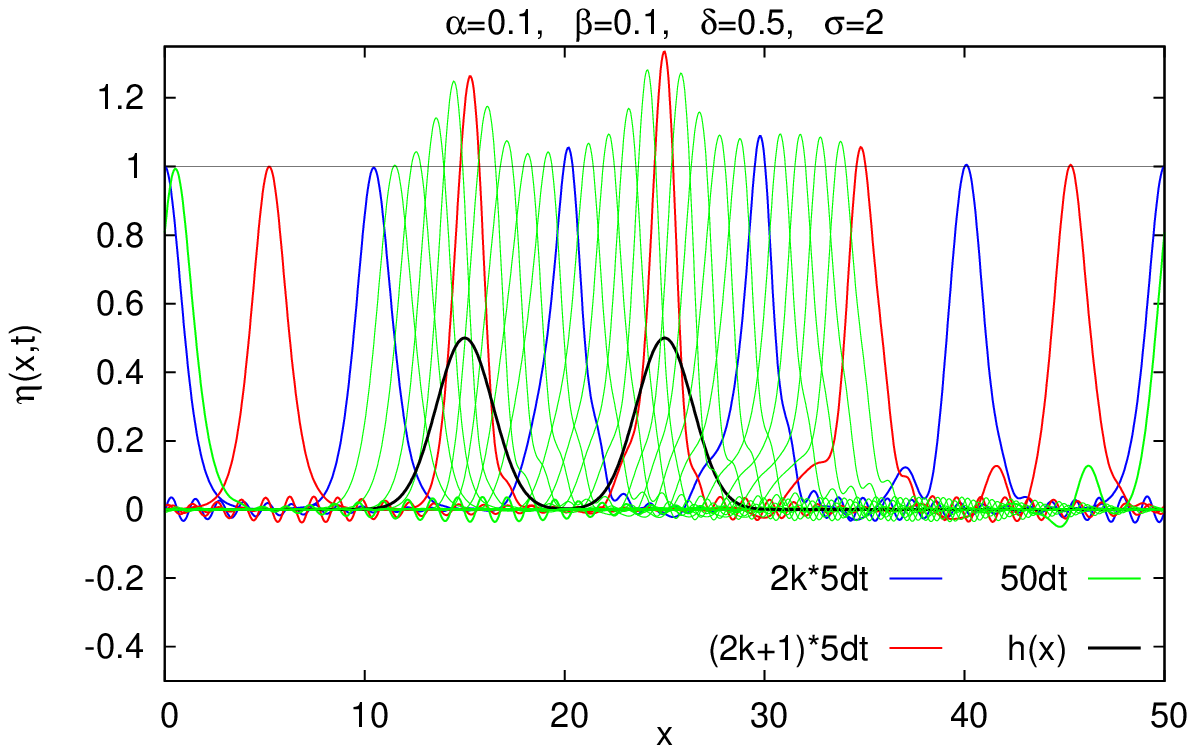}}
\caption{Motion of the soliton according to the equation (\ref{etaabd}) for a sequence of two bupms.} \label{F10}
\end{figure}\end{center}

In Figs.\ \ref{F9} and \ref{F10} the case of KdV soliton passing over two Gaussian wells is compared to the case of the soliton passing over two Gaussian bumps. In the former case one does not see substantial changes of the soliton's shape, only the amplitude oscillates (with corresponding oscillation of the width in the opposite phase), when the soliton is over the region of the obstacles.
In the case of two bumps, however, one can clearly see a distortion of the soliton andan emergence of the secondary wave (soliton?) moving slower than the main one.

It should be stressed that in all presented cases the volume of the wave was conserved with very high numerical accuracy (constant up to 10 digits).

\section{Summary}
The new nonlinear equation for shallow water wave problem in weakly nonlinear, dispersive and  long wavelength limit was derived. The equation contains terms originating from variable bottom under assumptions that the amplitude of the bottom function and its derivatives are relatively small. 
The method applied in the paper can be used for derivation of the wave equation
in more general cases, when some other small effects like surface tension, small viscosity etc.\ are taken into account. The numerical simulations exhibit the fact that KdV soliton persist its form even for substantial changes of the bottom.

 \begin{ack}
One of the authors (A.K.) thanks for support from "Stochastic Analysis Research Network", grant PIA-CONICYT-Scientific Research Ring \#1112.
\end{ack}

\end{document}